\documentclass[12pt]{article}

\usepackage{graphicx}
\usepackage{natbib}
\usepackage{amsbsy}
{\providecommand\boldsymbol[1]{\mbox{\boldmath $##1$}}}

\providecommand\bnabla{\boldsymbol{\nabla}}
\providecommand\bcdot{\boldsymbol{\cdot}}

\newcommand\Pran{\mbox{\textit{Pr}}} 

\newsavebox{\astrutbox}
\sbox{\astrutbox}{\rule[-5pt]{0pt}{20pt}}

\title{\Large \bf Finite difference solution of radiation on unsteady free convective magnetohydrodynamic flow past a vertical cylinder with heat and mass transfer}

\author{{\large M. N. Ismail$^1$, G. F. Mohamadien$^2$ and \"{U}mit D. G\"{o}ker$^3$}\hspace*{-1mm}
  \thanks{Email address for correspondence: deniz.goker@boun.edu.tr}
\\ {\normalsize $^1$ Astronomy Department, Faculty of Science, Al-Azhar University}\\ {\normalsize 11884, Cairo,
 Egypt}\\[1mm]
{\normalsize $^2$ National Research Institute of Astronomy and Geophysics, Helwan}\\ {\normalsize 11884, Cairo, Egypt}\\[1mm]
{\normalsize $^3$ Department of Physics, Bo\v{g}azi\c{c}i University, Bebek}\\ {\normalsize 34342, Istanbul, TURKEY}}
\date{}

\begin{document}
\maketitle

\begin{abstract}
We study effects of both magnetic field and radiation on unsteady free convective magnetohydrodynamic (MHD) flow under the influence of heat and mass transfer.
An implicit finite difference scheme of Crank-Nicolson method is used to analyze the results. The governing boundary layer equations along with the initial
and boundary conditions are casted into a dimensionless form which results in a more precise solution. We determine the change in velocity under the influence
of radiation and magnetic field with temperature and concentration taken into account. A code is constructed (FORTRAN 6.5) to obtain predictions following from
the numerical solution which have been compared with the existing information in the literature, and a very good agreement is obtained. The results obtained
through this study will be helpful in the prediction of flow and heat transfer in the presence of magnetic field and radiation.
\end{abstract}

\section{Introduction}
\label{sec-intro}
The flow problems occur in many engineering, technology and space sciences applications. From an industrial point of view, the radiation effect of free
convective magnetohydrodynamic (MHD) flow with heat transfer has become more important in recent years. Many processes in engineering phenomena occur
at high temperatures and the knowledge of radiation heat transfer has become very significant for the design of pertinent equipment at these temperatures.
The unsteady natural convection flow over a vertical cylinder with MHD flow has been given an extensive attention in literature. Several theoretical and
experimental investigations on heat and mass transfer characteristics, radiation effects and free convection have been carried out from the beginning of sixties.
Heat transfer problem from vertical circular cylinders to the surrounding initially quiescent fluids has first been analyzed by \citet{Gold64} for the transient
free convection. The effects of heat and mass transfer on the naturally convective flow along a vertical cylinder were studied by \citet{chen80}.
\citet{Raptis82} have reported the influence of magnetic field upon the steady free convective flow through a porous medium bounded by an infinite vertical plate
with constant section velocity where the plate temperature was assumed constant. By using harmonic analysis, \citet{Takhar91} have focused the effects of Hall
currents on MHD free convection boundary layer flow via a porous medium past through a plate. \citet{Velus92} have given a numerical solution for the transient
natural convection over heat by generation of vertical cylinders of various thermal capacities and radii. However, a fully implicit finite difference technique was
used to solve non-linear set of equations by \citet{Velus92}. \citet{Helmy98} studied the unsteady laminar free convection flow of an electrically conducting fluid
through a porous medium bounded by an infinite vertical plane surface of constant temperature. The radiation effect on heat transfer over a stretching surface has
first been studied by \citet{Elbash00}. Recently, \citet{Ghaly00} investigated the radiation effect on MHD free convective flow of a gas at a stretching surface
with a uniform free stream. \citet{Seddeek01} has studied thermal radiation and buoyancy effects on MHD free convective heat generated flow over an accelerating
permeable surface with temperature dependent viscosity. \citet{medic01} have studied the radiation effects on unsteady MHD free convection Hall current near and
infinite vertical porous plate.

However, no magnetic field and radiation working together on unsteady free convective flow along vertical cylinder under the influence of heat and mass transfer
has been reported. Due to this fact, the main objective of the present work is to investigate the problem of unsteady free convective MHD flow of an incompressible
viscous fluid passing through a vertical cylinder under the influence of a magnetic field and radiation. Thus, under the influence of radiation and magnetic field
we were able to achieve the important changes in velocity, temperature and concentration. For this purpose, results from an implicit finite difference scheme of
Crank-Nicolson method were used. In \S2.1, the governing equations are given. In \S2.2 the details of the numerical solutions are described and the main features
of the flow are presented. In \S3, the numerical results and discussions are presented. In \S4, the conclusions are given.

\section{Numerical database}
\label{sec-data}
\subsection{The governing equations}
\label{subsec-eqns}
In this section, the effect of radiation on unsteady free convective MHD flow is analyzed. The governing equations are solved more accurately than those in the
previous works on the subject. The numerical predictions have been compared with the existing information in the literature ($e.g.$ \citet{Pull08};
\citet{Deka11}) and a very good agreement is obtained. Axial coordinate $x$ is measured along the axis of the cylinder. The radial coordinate $r$ is measured
normal to the $x$-axis of the cylinder. We consider the MHD flow of a viscous incompressible fluid past through a vertical cylinder of radius $r_0$ with
transversely applied magnetic field. The radiating gas is said to be non-gray if its absorption coefficient is dependent on wavelength. The equation that described
the conservation of radiative transfer in a unit volume for all wavelengths is
\begin{equation}\label{radiat}
    \bnabla\bcdot\bar{\textbf{\textit{q}}}_r = \int\limits_0^\infty k_\lambda(T)\left(4\pi I_{b\lambda}(T) - G_\lambda\right)\,d\lambda
\end{equation}
where $\bar{\textbf{\textit{q}}}_r$ is the average heat flux of the radiation, $I_{b\lambda}$ is the spectral intensity for a black body, $k_\lambda$ is the
absorption coefficient, and the incident radiation $G_\lambda$ is defined as
\begin{equation}\label{G}
    G_\lambda = \int\limits_{\Omega=4\pi}I_{b\lambda}(\Omega)\,d\Omega
\end{equation}
where $\Omega$ is the solid angle. For an optically thin fluid exchanging radiation with an isothermal flat plate, according to Equation (\ref{G}) and Kirchhoff
law, the incident radiation is given by
\begin{equation}\label{G1}
   G_\lambda = 4\pi I_{b\lambda}(T_w) = 4e_{b\lambda}(T_w)
\end{equation}
where $T$ is the average temperature value of the porous plate. Equation (\ref{radiat}) is then reduced to
\begin{equation}\label{radiat1}
     \bnabla\bcdot\bar{\textbf{\textit{q}}}_r = \int\limits_0^\infty k_{\lambda w}(T)\bigl(e_{b\lambda}(T) - e_{b\lambda}(T_w)\bigr)\,d\lambda
\end{equation}
where $k_{\lambda w} = k_\lambda(T_w)$ is the mean absorption coefficient, $e_{b\lambda}$ is Planck's function, and $T$ is the temperature of the fluid in a
boundary layer. After expanding $e_{b\lambda}(T)$ and $k_{\lambda w}(T)$ in Taylor series around $T_w$ for a small parameter $(T-T_w)$, the Equation (\ref{radiat1})
becomes
\begin{equation}\label{radiat2}
   \bnabla\bcdot\bar{\textbf{\textit{q}}}_r = - 4\Gamma(T - T_w)
\end{equation}
where $\Gamma$ is a constant of integration. All fluid properties are considered to be constant except for the density variation which induces the buoyancy force.
The viscous dissipation is neglected in the energy equation. The governing equations are expressed in terms of conservation of mass, momentum, energy, mass
diffusion and the interaction of the flow with the magnetic field. We assumed that there is no chemical reaction between diffusing species and the fluid.
Initially, it is assumed that the fluid and the cylinder are of the same temperature $T'_\infty$ and also at same concentration $C'_\infty$. Then the surface
temperature and concentration of the cylinder are raised to a uniform temperature and concentration $T'_w$ and $C'_w$, respectively. Here, the subscripts $w$ and
$\infty$ indicate the conditions on the wall and free stream, respectively. It is further assumed that the interaction of the induced axial magnetic field with
the flow is considered to be negligible compared to the interaction of the applied magnetic field $B_0$. Then the flow is governed by the following system of
equations \\[1cm]
\begin{eqnarray}
   & & \frac{\partial(ru)}{\partial x} + \frac{\partial(rv)}{\partial r} = 0
   \label{mass} \\
   & & \frac{\partial u}{\partial t'} + u\frac{\partial u}{\partial x} + v\frac{\partial u}{\partial r}
   = gB(T' - T'_\infty) + gB^\ast(C' - C'_\infty)\nonumber \\
   & &\phantom{\frac{\partial u}{\partial t'} + u\frac{\partial u}{\partial x}\frac{v}{r}\frac{\partial}{\partial r} + }\hspace*{3pt}
   + \frac{v}{r}\frac{\partial}{\partial r}\left(r\frac{\partial u}{\partial r}\right)
   - \sigma\frac{B_0^2}{\pi\rho}u
   \label{momentum} \\
   & & \frac{\partial T'}{\partial t'} + u\frac{\partial T'}{\partial x} + v\frac{\partial T'}{\partial r} =
   \frac{\alpha}{r}\frac{\partial}{\partial r}\left(r\frac{\partial T'}{\partial r}\right) - 4\Gamma(T - T_w)
   \label{energy} \\
   & & \frac{\partial C'}{\partial t'} + u\frac{\partial C'}{\partial x} + v\frac{\partial C'}{\partial r} =
   \frac{D}{r}\frac{\partial}{\partial r}\left(r\frac{\partial C'}{\partial r}\right)
   \label{diff}
\end{eqnarray}
The equations above indicate the continuity, momentum and energy conservation, and mass diffusion equations, respectively. Here, $u$, $v$, $t'$, $g$, $B$, $T'$,
$B^\ast$, $C'$, $\sigma$, $\rho$, $\alpha$ and $D$ represent the $x$-component of velocity, the radial component of velocity, time, acceleration due to gravity,
magnetic field (for temperature), temperature, modified magnetic field (for concentration), concentration, Stefan-Boltzmann constant, density, thermal diffusivity
of fluid and mass diffusion coefficient, respectively. The initial and boundary conditions are given by\\
\begin{equation}\label{cond}
  \left.  \begin{array}{lcl}
      t'\le 0: & u = 0, & v = 0, \\
      T' = T'_\infty, & C' = C'_\infty & \textrm{for all}\; x \;\textrm{and}\; r \\
       &  &  \\
      t'> 0: & u = 0, & v = 0, \\
      T' = T'_w, & C' = C'_w & \textrm{at}\; r = r_0
    \end{array} \right\}
\end{equation}\\

We introduce the following dimensionless quantities\\
\begin{equation}\label{nondim}
    \left.
    \begin{array}{lcl}
      X = Gr^{-1} x/r_0, & R = r/r_0, & U = ur_0Gr^{-1}/v, \\[2mm]
      V = vr_0/v, & t = vt'/r_0^2,  & Sc=v/D, \\[2mm]
      F=4\Gamma r_0^2/(vG\sqrt{r}), & N=Gr^\ast/(Gr), & M=\sigma B_0^2r_0^2/(\rho v), \\[2mm]
      Pr=v/\alpha, & \alpha=k/(\rho c_p), & \theta = {\displaystyle\frac{T' - T'_\infty}{T'_w - T'_\infty}}, \\[4mm]
      C = {\displaystyle\frac{C'-C'_\infty}{C'_w-C'_\infty}}, &
      Gr={\displaystyle\frac{gBr_0^3(T'_w-T'_\infty)}{v^2}}, & Gr^\ast={\displaystyle\frac{gB^\ast r_0^3(C'_w-C'_\infty)}{v^2}}.
    \end{array}
    \right\}
\end{equation}\\
Here $X$, $R$, $U$, $V$, $t$, $N$, $M$, \Pran, $Gr$, $Gr^\ast$, \textit{$\Theta$}, $Sc$, $k$, $c_p$ and $F$ are the dimensionless variables such as spatial coordinate, radial coordinate, $x$-component of velocity, radial velocity, time, the combined buoyancy ratio parameter, magnetic field parameter, Prandtl number, Grashof number (for temperature), modified Grashof number (for concentration), temperature, Schmidt number, thermal conductivity, specific heat at constant pressure and the radiation parameter, respectively. The Equations (\ref{mass})--(\ref{diff}) are reduced to the following form
\begin{eqnarray}
   & & \frac{\partial(RU)}{\partial X} + \frac{\partial(RV)}{\partial R} = 0
   \label{mass1} \\
   & & \frac{\partial U}{\partial t} + U\frac{\partial U}{\partial X} + V\frac{\partial U}{\partial R}
   = \theta + NC +\frac{1}{R}\left(R\frac{\partial U}{\partial R}\right) - MU
     \label{momentum1} \\
   & & \frac{\partial\theta}{\partial t} + U\frac{\partial\theta}{\partial X} + V\frac{\partial\theta}{\partial R} =
   \frac{1}{Pr R}\frac{\partial}{\partial R}\left(R\frac{\partial\theta}{\partial R}\right) +F(\theta-1)
   \label{energy1} \\
   & & \frac{\partial C}{\partial t} + U\frac{\partial C}{\partial X} + V\frac{\partial C}{\partial R} =
  \frac{1}{Sc R}\frac{\partial}{\partial R}\left(R\frac{\partial\theta}{\partial R}\right)
   \label{diff1}
\end{eqnarray}\\
The corresponding initial and boundary conditions expressed in the non-dimensional quantities are given by:\\
\begin{equation}\label{cond1}
  \left.  \begin{array}{lccl}
      t'\le 0: & U = 0, & V = 0, & \\
     & \theta=0, & C=0, &\textrm{for all}\; X \;\textrm{and}\; R \\
       &  &  & \\
      t'> 0: & U = 0, & V = 0, & \\
       & \theta=1, & C = 1, & \phantom{C=0}\hspace*{3mm}\textrm{at}\; R = 1 \\
       & U=0, & \theta=0, & C=0\hspace*{3mm}\textrm{at}\; X=0 \\
       & U\rightarrow 0, & \theta\rightarrow 0, & C\rightarrow 0\hspace*{2mm} \textrm{as}\; R\rightarrow \infty
    \end{array} \right\}
\end{equation}
\\
\vspace*{1mm}

\subsection{Numerical technique}
\label{subsec-numeric}
The unsteady, non-linear, coupled partial differential equations (\ref{mass1})-(\ref{diff1}) with the initial and boundary conditions (\ref{cond1}) are solved by
employing a finite difference scheme of Crank-Nicholson method which was discussed by many authors ($e.g.$ \citet{Pull08}; \citet{Deka11}). The Crank-Nicholson
method is a finite difference method (\citet{crank47}) and it solves the heat equation and similar partial differential equations numerically. For the heat equation,
the method consider the normalized heat equation in one dimension with homogeneous Dirichlet boundary conditions. This method is based on central difference in
space and gives the second-order convergence in time. Equivalently, it is the average of forward and backward Euler schemes in time. In the applications to large
time steps or high spatial resolution, the less accurate backward Euler method can be used. The Euler method is both stable and independent from oscillations
(\citet{crank47}). The partial differential equations (\ref{mass1})--(\ref{diff1}) are converted into finite difference equations:\\
\begin{eqnarray}
   & & \frac{1}{4 \Delta x}(U^{n+1}_{i,j}-U^{n+1}_{i-1,j}+U^{n+1}_{i,j-1}-U^{n+1}_{i-1,j-1}+U^{n}_{i,j}-U^{n}_{i-1,j}+U^{n}_{i,j-1}-U^{n}_{i-1,j-1})\nonumber \\
   & & + \frac{1}{2 \Delta R}(V^{n+1}_{i,j}-V^{n+1}_{i,j-1}+V^{n}_{i,j}-V^{n}_{i,j-1})=0
     \label{fini1} \\[7mm]
   & & \phantom{\frac{1}{4 \Delta t}(U^{n+1}_{i,j}-U^{n}_{i,j})+\frac{1}{2 \Delta x}U^{n}_{i,j}(U^{n+1}_{i,j}-U^{n}_{i-1,j}+U^{n}_{i,j}-U^{n}_{i-1,j})}\nonumber
   \\[-10mm]
   & & \frac{1}{\Delta t}(U^{n+1}_{i,j}-U^{n}_{i,j})+\frac{1}{2 \Delta x}U^{n}_{i,j}(U^{n+1}_{i,j}-U^{n+1}_{i-1,j}+U^{n}_{i,j}-U^{n}_{i-1,j})\nonumber \\
   & & +\frac{1}{4 \Delta R}V^{n+1}_{i,j}(U^{n+1}_{i,j+1}-U^{n+1}_{i,j-1}+U^{n}_{i,j+1}-U^{n}_{i,j-1})\nonumber \\
   & & = \frac{1}{2}(\theta^{n+1}_{i,j}+\theta^{n}_{i,j})-\frac{M}{2}(U^{n+1}_{i,j}+U^{n}_{i,j})+\frac{N}{2}(C^{n+1}_{i,j}+C^{n}_{i,j})\nonumber\\
   & & +\frac{1}{2 (\Delta R)^{2}}(U^{n+1}_{i,j-1}-2U^{n+1}_{i,j}+U^{n+1}_{i,j+1}+U^{n}_{i,j-1}-2U^{n}_{i,j}+U^{n}_{i,j+1})
     \label{fini2} \\[4mm]
   & & \frac{1}{\Delta t}(\theta^{n+1}_{i,j}-\theta^{n}_{i,j})+\frac{1}{2 \Delta x}U^{n}_{i,j}(\theta^{n+1}_{i,j}-\theta^{n+1}_{i-1,j}+\theta^{n}_{i,j}
   -\theta^{n}_{i-1,j})\nonumber\\
   & & +\frac{1}{4 \Delta R}V^{n+1}_{i,j}(\theta^{n+1}_{i,j+1}-\theta^{n+1}_{i,j-1}+\theta^{n}_{i,j+1}-\theta^{n}_{i,j-1})\nonumber \\
   & &  = \frac{1}{2 (\Delta R)^{2} Pr}(\theta^{n+1}_{i,j-1}-2\theta^{n+1}_{i,j}+\theta^{n+1}_{i,j+1}+3\theta^{n}_{i,j+1}+\theta^{n}_{i,j-1}-2\theta^{n}_{i,j})
   \nonumber\\
   & & - F(\frac{1}{2}(\theta^{n+1}_{i,j}+\theta^{n}_{i,j}-1))
     \label{fini3} \\[4mm]
   & & \frac{1}{\Delta t}(C^{n+1}_{i,j}-C^{n}_{i,j})+\frac{1}{2 \Delta x}U^{n}_{i,j}(C^{n+1}_{i,j}-C^{n+1}_{i-1,j}+C^{n}_{i,j}-C^{n}_{i-1,j})\nonumber\\
   & & +\frac{1}{4 \Delta R}V^{n+1}_{i,j}(C^{n+1}_{i,j+1}-C^{n+1}_{i,j-1}+C^{n}_{i,j+1}-C^{n}_{i,j-1})\nonumber \\
   & & = \frac{1}{2 Sc (\Delta R)^{2}}(-C^{n+1}_{i,j}-C^{n+1}_{i,j+1}+2C^{n}_{i,j+1}-2C^{n}_{i,j})
     \label{fini4}
\end{eqnarray}

The code is constructed by using FORTRAN 6.5 to obtain the numerical solutions of the finite difference equations which were obtained from the Crank-Nicolson method.

\section{Results and discussions}
\label{sec:numerical_results}
\begin{figure}
  \centerline{\includegraphics{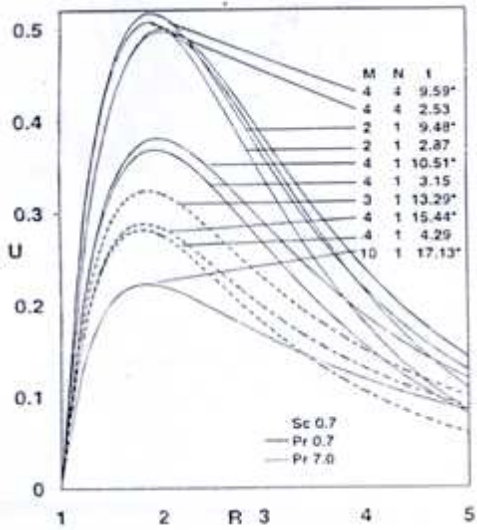}}
  \caption{The velocity profiles for $M$ and $N$ at different times with the effects of $\Pran$ numbers, $Pr = 0.7, 7.0$ and $Sc$ number, $Sc = 0.7$.
  Transient velocity profiles are indicated at $X=1$ and $F=0$. The value of $t$ with star (*) symbol denotes the time taken to reach steady state. The $R$
  direction denotes the radial coordinate and $U$ direction denotes $x$-component of the velocity. Here, $M$ and $N$ indicate the magnetic field parameter and
  the combined buoyancy ratio parameter, respectively.}
\label{fig:1}
\end{figure}

\begin{figure}
  \centerline{\includegraphics[scale=.9]{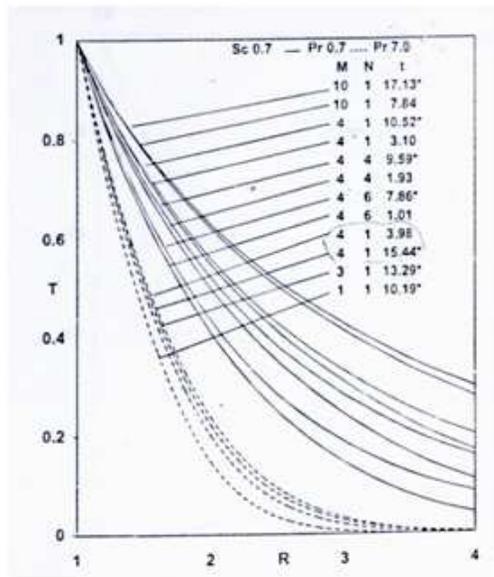}}
  \caption{The temperature profiles for $M$ and $N$ at different times with the effects of $\Pran$ numbers, $Pr = 0.7, 7.0$ and $Sc$ number, $Sc = 0.7$. Transient
  temperature profiles are indicated at $X=1$ and $F=0$. The value of $t$ with star (*) symbol denotes the time taken to reach steady state. The $R$ direction
  denotes the radial coordinate and $T$ direction denotes the temperature.}
\label{fig:2}
\end{figure}

\begin{figure}
  \centerline{\includegraphics{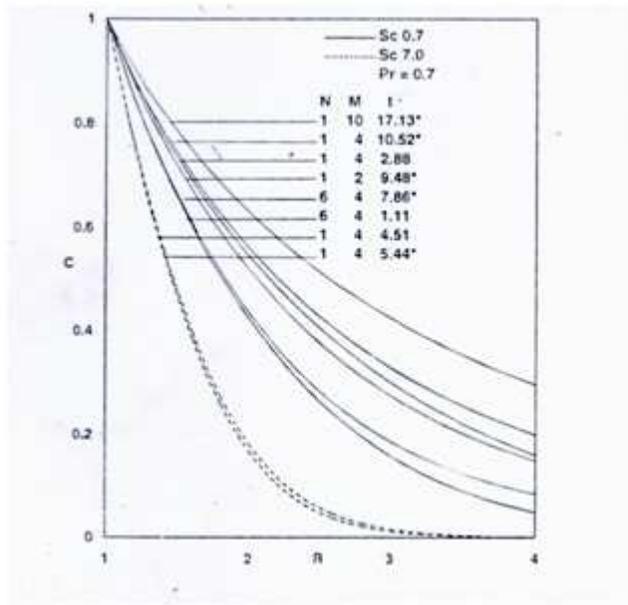}}
  \caption{The concentration profiles for $M$ and $N$ at different times with the effects of $\Pran$ numbers, $Pr = 0.7, 7.0$ and $Sc$ number, $Sc = 0.7$.
  Transient concentration profiles at $X=1$ and $F=0$. The value of $t$ with star (*) symbol denotes the time taken to reach steady state. The $R$ direction
  denotes the radial coordinate and $C$ direction denotes concentration.}
\label{fig:3}
\end{figure}

In order to get an insight into the physics of the problem, we present the results of numerical computations of velocity, temperature and concentration for
different values of Prandtl number $\Pran$, Schmidt number $Sc$, magnetic field parameter $M$, combined buoyancy ratio parameter $N$, time $t$ and radiation
parameter $F$. $\Pran$ physically relates the relative thickness of hydrodynamic and thermal boundary layers while $Sc$ relates the relative thickness of
hydrodynamic and concentration boundary layers (\cite{Deka11}).
Velocity profiles represented by Fig.\ref{fig:1} show the effect of $\Pran$ and $Sc$ numbers for different values of time near the cylinder at $X=1$. Effects of
$\Pran$ and $Sc$ numbers on the transient temperature profiles are given in Fig.\ref{fig:2} and transient concentration profiles in Fig.\ref{fig:3} with respect
to time, near to cylinder at $X=1$. The effect of $\Pran$ number is very important in temperature field especially. The radiation effect had not been taken into
account in the first three figures.

\begin{figure}
  \centerline{\includegraphics[scale=.9]{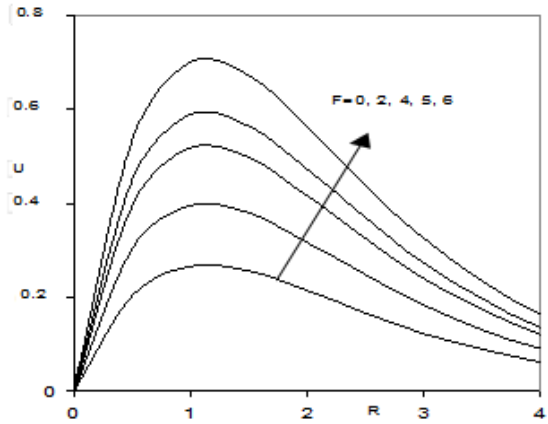}}
  \caption{Transient velocity profiles at $X=1$, $M=4$, $N=1$, $t=1$, $Sc=0.7$ and $Pr=0.7$. Different plots show different radiation effects for the velocity
  profiles. From the bottom profile to the top, the values of $0, 2, 4, 5$ and $6$ indicate the changes of temperature without and with radiation, respectively.
  Increasing radiation values have increased the slopes of velocity profiles. The $R$ direction denotes the radial coordinate and $U$ direction denotes
  $x$-component of the velocity.}
\label{fig:4}
\end{figure}

\begin{figure}
  \centerline{\includegraphics[scale=.8]{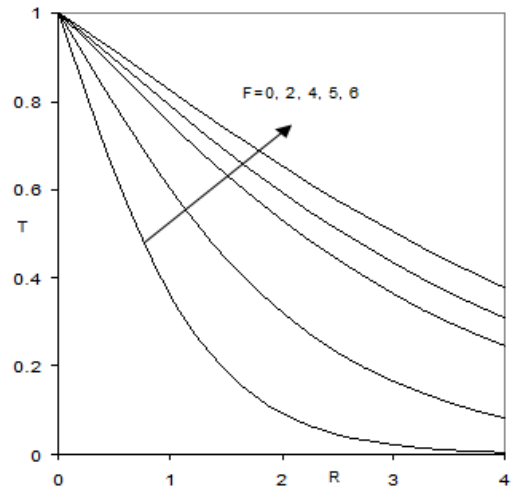}}
  \caption{Transient temperature profiles at $X=1$, $M=4$, $N=1$, $t=1$, $Sc=0.7$ and $Pr=0.7$. Different plots show different radiation effects for the
  temperature profiles. From the bottom profile to the top, the values of $0, 2, 4, 5$ and $6$ indicate the changes of temperature without and with radiation,
  respectively. In the absence of radiation, a very deep slope stands out. The $R$ direction denotes the radial coordinate and $T$ direction denotes the
  temperature.}
\label{fig:5}
\end{figure}

\begin{figure}
  \centerline{\includegraphics[scale=.7]{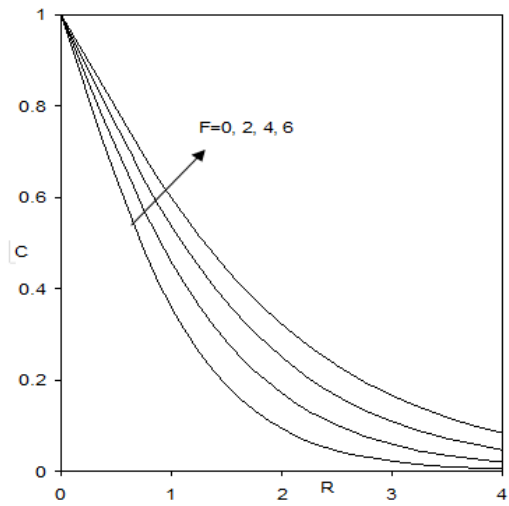}}
  \caption{Transient concentration profiles at $X=1$, $M=4$, $N=1$, $t=1$, $Sc=0.7$ and $Pr=0.7$. Different plots show different radiation effects for the
  concentration profiles. From the bottom profile to the top, the values of $0, 2, 4$ and $6$ indicate the changes of concentration without and with radiation,
  respectively. The $R$ direction denotes the radial coordinate and $C$ direction denotes the concentration.}
\label{fig:6}
\end{figure}

The transient profiles of velocity, temperature and concentration in the presence of radiation parameter $F$ are indicated in Fig.\ref{fig:4}, Fig.\ref{fig:5} and
Fig.\ref{fig:6} with respect to spatial coordinate $X$, magnetic field parameter $M$, combined buoyancy ratio parameter $N$, time $t$, Prandtl $\Pran$ and Schmidt
$Sc$ numbers. All these numbers are in the dimensionless form as indicated above. To study the behavior of the velocity, temperature and concentration profiles,
a code using FORTRAN (version 6.5) is constructed to solve Equations (\ref{fini1})-(\ref{fini4}). Following these results from the simulation for an unsteady free
convective MHD flow past a vertical cylinder with heat and mass transfer, the following general points can be made: \\

i) In Fig.\ref{fig:1}, the velocity increases with time and reaches the steady state in a certain lapse of time. Temporal maximum is not observed. The $R$ direction
denotes the radial coordinate and $U$ direction denotes the $x$-component of the velocity. The Schmidt number $Sc$ is taken as 0.7, while the Prandtl numbers
$\Pran$ are indicated as 0.7 and 7.0 with solid and dashed lines, respectively. Hence, the transient velocity decreases with increase in $\Pran$ number. When $N$,
the combined the ratio parameter, increases the velocity increases. The velocity profile decreases with increasing magnetic field parameter $M$.  Time taken to
reach the steady state increases with increasing $M$.

ii) In Fig.\ref{fig:2}, higher $\Pran$ values show higher heat transfer rates. The $R$ direction denotes the radial coordinate and $U$ direction denotes the
$x$-component of the velocity without the effect of radiation. The transient temperature profiles increase with decrease in the value of $\Pran$ but increases with
increasing time. The higher values of time $t$ show the steady state conditions. The Prandtl numbers $\Pran$ are indicated as 0.7 and 7.0 with solid and dashed
lines, respectively. The temperature profiles increase with increasing magnetic field parameter $M$ but it will have decreasing slopes with increasing temperatures
in the limit of the combined buoyancy ratio parameter $N\rightarrow1$.

iii) Unsteady concentration profiles are shown in Fig.\ref{fig:3}. The $R$ direction denotes the radial coordinate and $U$ direction denotes the $x$-component of
the velocity. We hold the Prandtl number $Pr$ steady and change the Schmidt number $Sc$. When $Sc$ increases, the mass transfer rate increases. The solid and
dashed lines represent different values of $Sc$ numbers as 0.7 and 7.0, respectively. The transient concentration decreases with the increase in value of $Sc$ and
increases with increasing time $t$ and reached the steady state for larger time $t$. The concentration profiles increase with increasing magnetic field parameter
$M$ while it does not show larger effect in the limit of the combined buoyancy ratio parameter $N\rightarrow1$. In the case of increasing $N$, the concentration
profiles are observed to have decreasing slopes with increasing concentration.

iv) The effect of radiation parameter $F$ on the velocity, temperature and concentration is shown in Fig.\ref{fig:4}, Fig.\ref{fig:5} and Fig.\ref{fig:6},
respectively. It is observed that the velocity, temperature and concentration increase as the radiation parameter $F$ increases. This result qualitatively agrees
with expectations since the effects of radiation and surface temperature are to increase the rates of energy transport to the fluid, thus increasing the temperature
of the fluid.

v) In Fig.\ref{fig:4}, the variation of velocity profiles in the presence of radiation parameter are indicated. The $R$ direction denotes the radial coordinate
and $U$ direction denotes the $x$-component of the velocity without ($F=0$) and with ($F=2, 4, 5, 6$) the effect of radiation. The velocity profiles (and slopes)
increase with increasing radiation values. The velocity is related to the rate of energy transport of the fluid. As long as the radiation increases the energy
transport rate, the velocity will also be increased.

vi) There is a direct relationship between temperature and radiation. It can clearly be seen in Fig.\ref{fig:5}. The $R$ direction denotes the radial coordinate
and $T$ direction denotes the temperature without ($F=0$) and with ($F=2, 4, 5, 6$) the effect of radiation. Therefore, the temperature of the fluid increases with
increasing radiation and increasing the radiation will stop the sudden decline in the temperature as seen in the slope of the bottom temperature profile.

vii) The effect of radiation for the concentration profiles is seen in Fig.\ref{fig:6}. The $R$ direction denotes the radial coordinate and $C$ direction denotes
the concentration without ($F=0$) and with ($F=2, 4, 6$) the effect of radiation. Although there was not a very big difference in concentration, the noticeable
increase in the radial coordinate is very obvious. Mostly, the radiation effects the velocity and temperature profiles. The less deep but similar profiles can also
be seen in concentration.


\section{Conclusions}
\label{sec:conclusions}

In this work, we have made a numerical analysis for unsteady free convective MHD flow past a vertical cylinder with heat and mass transfer under the effect
of magnetic field and radiation by employing finite difference scheme of Crack-Nicolson method. For the first time, we included in our analysis the magnetic
field and radiation parameters together. The velocity, temperature and concentration profiles are drawn for different values of parameters such as
$\Pran$, $Sc$, $M$, $N$, $t$ and $F$ to study the behavior of velocity, temperature and concentration of the flow starting from the initial radial
coordinate $R_0 = 4$. Present results have been compared with the available results in literature and are found to be in a good agreement.
Later, we have examined the effect of radiation for the same profiles. The radiation will be more effective on the profiles of velocity, temperature
and concentration. Increasing the radiation implies increasing of velocity and temperature in the flow. The flow problems at high operating temperatures
where the radiation effects are quite significant occur in many engineering applications, science and technology. The effects of free convective heat flow
play an important role in agriculture, petroleum, industries, geological formations. The results obtained from this study will be helpful in the prediction
of flow and heat transfer.

\section{Acknowledgement}
The authors would like to thank Professor Mikhail Sheftel at Bo\u{g}azi\c{c}i University, Department of Physics for fruitful discussions and valuable comments.

\end{document}